\documentclass[journal=jacsat,manuscript=article]{achemso}

\usepackage[version=3]{mhchem} 

\usepackage{braket}
\usepackage{verbatim}

\author{Jeffrey Rable}
\email{j.rable@northeastern.edu}
\altaffiliation{Corresponding Author}
\affiliation[NU Physics]{Department of Physics, Northeastern University, Boston, MA 02115}
\alsoaffiliation[QMSI]{Quantum Materials and Sensing Institute, Northeastern University, Burlington, MA, 01803}
\author{Jyotirmay Dwivedi}
\affiliation[psu]{Department of Physics, Pennsylvania State University, State College, PA, 16802}
\author{Nitin Samarth}
\affiliation[psu]{Department of Physics, Pennsylvania State University, State College, PA, 16802}
\author{Paul Stevenson}
\affiliation[NU Physics]{Department of Physics, Northeastern University, Boston, MA 02115}
\alsoaffiliation[QMSI]{Quantum Materials and Sensing Institute, Northeastern University, Burlington, MA, 01803}
\author{Arun Bansil}
\affiliation[NU Physics]{Department of Physics, Northeastern University, Boston, MA 02115}
\alsoaffiliation[QMSI]{Quantum Materials and Sensing Institute, Northeastern University, Burlington, MA, 01803}
\author{Swastik Kar}
\affiliation[NU Physics]{Department of Physics, Northeastern University, Boston, MA 02115}
\alsoaffiliation[QMSI]{Quantum Materials and Sensing Institute, Northeastern University, Burlington, MA, 01803}
\alsoaffiliation[NU ChemE]{Department of Chemical Engineering, Northeastern University, Boston, MA 02115}
\email{s.kar@northeastern.edu}
\altaffiliation{Corresponding Author}

\title[]
  {Flux channeling induced nano-confinement and enhancement of microwaves imaged by Rabi oscillation mapping}

\abbreviations{NV,SNVM,Py,RF,MW,FMR,PL}

\begin{document}

\begin{abstract}
With rapid advances in qubit technologies, techniques for localizing, modulating, and measuring RF fields and their impact on qubit performance are of the utmost importance. Here, we demonstrate that flux-channeling from a permalloy nanowire can be used to achieve localized spatial modulation of an RF field and that the modulated field can be mapped with high resolution using the Rabi oscillations of an NV center. Rabi maps reveal {\raise.17ex\hbox{$\scriptstyle\mathtt{\sim}$}}100 mm wavelength microwaves concentrated in sub-300 nm-scale regions with up to {\raise.17ex\hbox{$\scriptstyle\mathtt{\sim}$}}16$\times$ power enhancement. This modulation is robust over a 20 dBm power range and has no adverse impact on NV  $T_2$ coherence time. Micromagnetic simulations confirm that the modulated field results from the nanowire’s stray field through its constructive/destructive interference with the incident RF field. Our findings provide a new pathway for controlling qubits, amplifying RF signals, and mapping local fields in various on-chip RF technologies.
\end{abstract}


The advent of practical quantum technologies has the potential to revolutionize fields ranging from computing to biosensing and spintronics.  However, as these technologies evolve beyond the proof-of-concept stage, scalable new hardware capable of generating large, localized GHz-frequency microwave magnetic fields to control the underlying qubits are required. The simple approach to driving spin dynamics - using Oersted fields generated by a microwave antenna - requires high power levels to produce sizable field amplitudes and their nonlocal nature becomes problematic in the nanoscale regime. Ferromagnetic dynamics have also been used to control Nitrogen-Vacancy (NV) spin defects in diamond, relying on coupling to magnons\cite{Andrich2017,Wang2020,Candido2020,Fukami2021,Simon2022,trifunovic_high-efficiency_2015} and proximal magnetic textures\cite{Wolf2016}. However, magnon-based techniques require tuning of magnetic fields to bring the NV transitions into resonance with magnon modes, while the latter approach suffers from problems associated with undesirable texture dynamics\cite{Badea2018,Badea2023}. 

Flux channeling via soft ferromagnets offers a promising alternative for increasing local field amplitudes. It is largely unexplored in the nanoscale and GHz regimes, although it has been used in NV center-based sensing of macroscopic, DC-kHz fields\cite{Xie2023, Fescenko2020} and in the engineering of mesoscopic transformers\cite{Yun2004,El-Ghazaly2017,Kerner2004}. Local amplification of fields in flux-channeling occurs when the relative permeability of a ferromagnetic material exceeds unity; this occurs up to the ferromagnetic resonance (FMR) frequency of the material, which can be increased via patterning\cite{El-Ghazaly2017}. 

The NV center, being a point defect, has a spatial resolution only limited by the NV-sample distance in a scanning system. It can be used to perform measurements from the kHz to GHz range with extremely high sensitivity\cite{Casola2018, Broadway2020}, down to 100's of $nT/\sqrt{Hz}$\cite{appel_nanoscale_2015} for resonant microwave fields, which makes the technique ideal for characterizing efficiency and losses in nanoscale RF devices. Unlike lower resolution techniques, scanning NV can probe the sources of loss, such as the localized losses driven by magnetic texture dynamics. Moreoever, the NV qubit allows the characterization of both the fields emitted by the RF devices and the impact the devices could have on the coherence of the defect spin, which are both valuable attributes for translation to applications in quantum information science. 

Here, we demonstrate the use of magnetic flux channeling in a 20 nm thick Permalloy (Ni-Fe alloy) soft ferromagnetic nanowire to locally modulate a microwave field at an NV center and show how spin state control can be enhanced in this way. For this purpose, we measured Rabi oscillations in a scanning NV magnetometer (SNVM), where the measured Rabi frequency varies linearly with the amplitude of the microwave driving field. A quantitative map of the field modulation is thus obtained such that $\Omega_R = \gamma_{\text{NV}} B_{\text{MW}}$, where $\Omega_R$ is the Rabi frequency, $\gamma_{\text{NV}}$ is the NV center's gyromagnetic ratio, and $B_{\text{MW}}$ is the amplitude of the driving field perpendicular to the NV center's quantization axis\cite{appel_nanoscale_2015}. Even at 84 nm sample-NV separation, we found maximum field enhancement to be 2.35$\times$, which corresponds to a 6-fold reduction in the needed microwave driving power at the highest power tested. In principle, this enhancement could be further improved through a reduction in the NV-sample distance and optimization of both the nanowire geometry and the microwave driving setup. 


\begin{figure} 
	\centering
	\includegraphics[width=\textwidth]{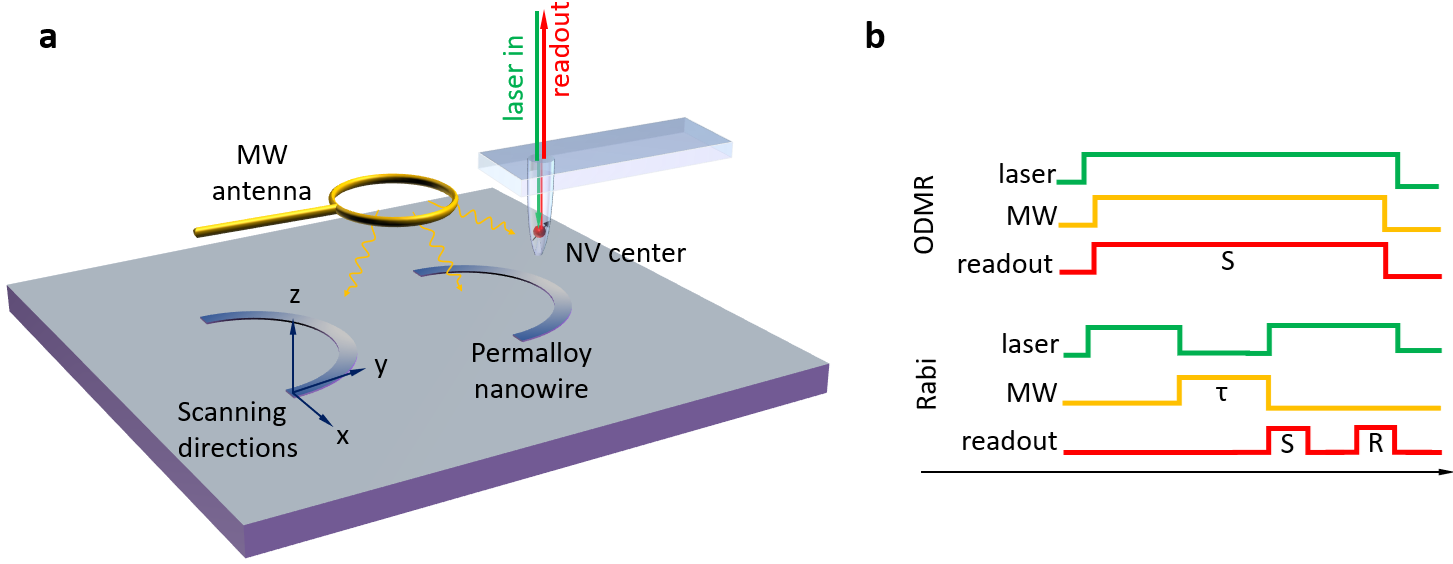} 
	\caption{A schematic of the experimental setup showcasing the measurement apparatus and pulse sequences. (a) Our scanning NV magnetometer positioned over our samples. A diamond scanning-probe tip is rastered across the sample, a 515 nm laser is used to polarize and read-out the NV spin state, and a gold wire loop acts as a microwave (MW) antenna to manipulate the spin. (b) Pulse sequences for ODMR and Rabi oscillation measurements. In the ODMR measurement, the MW frequency is swept while the 515 nm laser illuminates the diamond and the resulting photoluminescence (PL) is collected; the PL decreases when the NV $\ket{0} \rightarrow \ket{1}$ transition is driven by the microwave antenna. In a Rabi oscillation measurement, the spin is polarized into the $\ket{0}$ state using a 515 nm laser, and then a resonant microwave field driving the $\ket{0} \rightarrow \ket{1}$ transition is applied for a variable period of time $\tau$. Afterwards, the photoluminescence is read out, and a reference measurement without microwaves is done to measure the fluorescence of the brighter $\ket{0}$ state. This results is an optical contrast, given by the measured signal divided by the reference, which allows us to determine the spin state.}
	\label{fig:Cartoon} 
\end{figure}

Fig.~\ref{fig:Cartoon} provides a schematic of our experimental setup and measurement schemes. Flux channels were engineered using lithographically patterned semicircular (350 nm wide, 20 nm thick) Permalloy nanowires on a silicon substrate. The nanowires have a 5 µm diameter and a central, 140 nm wide half-circle notch defect. These wires are good candidates for flux channeling measurements with NV centers because (i) their predicted FMR frequency, which is over 7 GHz at single mT bias fields\cite{rable_off-resonant_2023}, is far above the NV’s 2.87 GHz zero-field transition frequency, and (ii) their relative thinness, which reduces losses from eddy currents. Two identical nanowires were fabricated simultaneously on two different chips to check reproducibility of the results.

Imaging of the samples was carried out with an SNVM instrument, shown in Figure~\ref{fig:Cartoon}(a), using optical excitation/readout combined with a MW antenna to drive both the NV spin and magnetic dynamics. Our scans combine measurements of topography, optically detected magnetic resonance (ODMR), and Rabi oscillations. The excitation pulses used are shown in Figure~\ref{fig:Cartoon}(b). In this way, we obtain quantitative information about the sample's stray field and its MW dynamics.

\begin{figure} 
	\centering
	\includegraphics[width=\textwidth]{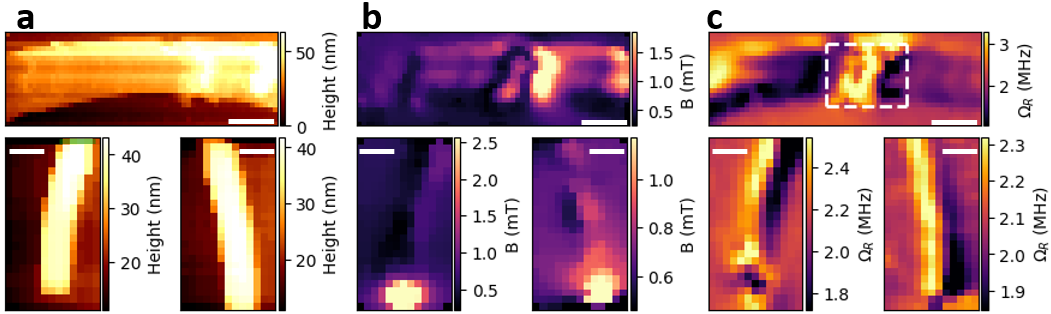}
	\caption{Results of Rabi oscillation scans, showing sample topography, stray field, and Rabi frequency. All scale bars represent 500 nm. (a) AFM topography map of the mid-section, the left end, and the right end of the nanowire, showing featureless, flat nanowire surfaces, with some height irregularities on the right side of the mid-section of the wire (b) ODMR measurements of the stray field from the wire at 139 nm NV-sample separation. The central area shows a large stray field from a domain wall, while the ends show large stray fields at the tips, where the in-plane magnetization points out of the sample. Other irregular features can be attributed to defects and grains in the wire. (c) Rabi maps, taken at the same height, showing localized regions of RF field enhancement and suppression. The data in the white dashed box in the mid-section is dominated by the presence of a domain wall, which resulted in Rabi oscillation measurements that yielded noisy data. This is likely due to Rabi frequencies lying outside our measurement range or poor optical contrast from large off-axis stray fields.}
	\label{fig:scans} 
\end{figure}

Figure 2 showcases the topographical, stray field, and Rabi maps measured in the x-y direction. We focused on the mid-section and the two ends of the permalloy nanowire structure to showcase RF modulations. Figure ~\ref{fig:scans}(a) shows the topography maps, which confirm featureless, flat surfaces with some irregularities on the right-side of the mid-section that we attribute to particulates on the sample. Figure ~\ref{fig:scans}(b) shows ODMR maps, all taken 139 nm above the three regions shown in Fig~\ref{fig:scans}(a). Notably, we found variations in the stray magnetic fields from the nanowire even in the topographically flat regions of the nanowires that were not correlated with the surface irregularities seen in Fig.~\ref{fig:scans}(a). The large central variation can be attributed to a domain wall pinned by the notch defect in the wire design, while other differences can be attributed to defects or grains in the wire. As expected, the ends of the nanowires show a large stray field from the magnetization of the wire, which is in-plane and oriented tangentially along the wire.  Figure ~\ref{fig:scans}(c) shows Rabi maps taken at the same height (139 nm), showing localized regions of strong RF field enhancement and suppression. Data in the boxed area (white dashed lines) in the mid-section is not relevant because of the presence of a domain wall, which made accurate measurements infeasible with our chosen experimental parameters. This could be the result of the large domain wall stray field\cite{Tetienne2015} reducing NV optical contrast, localized noise preventing coherent Rabi oscillations, or NV-domain wall interactions resulting in oscillations too fast or slow for our selected Rabi oscillation parameters. Outside of the boxed area, however, all three Rabi maps demonstrate a clear enhancement of the Rabi frequency along the left edges of the wire and an accompanying suppression on the right edges, indicating a proportional amplification and suppression of the applied microwave field. Approaching the center of the wire, we can further see the wire-orientation dependence of this effect: where the curvature is more prominent, we begin to see modulation of the Rabi frequency, whereas areas closer to the center of the wire show only a mild suppression or highly localized effects from a domain wall. Note that, outside the area in the immediate vicinity of the domain wall, the modulations of RF fields were strictly determined by the shape and orientation of the nanowire (relative to the direction of incidence of the RF source). This result is very promising, as it indicates that optimizing the geometry (size, shape, height) as well as the magnetic material could potentially lead to substantially higher modulation contrasts and localization of RF fields. Furthermore, it is quite remarkable to see that even for the very simple geometry of the nanowire, the field modulation remains tightly localized to an area on the order a few hundred nm, making it highly attractive for transfer of microwave power into nanoscale qubits and other devices.

We replicated our results on another wire with a second tip by performing height and power dependent measurements. The second nanowire was fabricated at the same time as the first wire, using the same design but a different substrate, and the second diamond scanning probe had a baseline NV-sample separation of 84 nm. These changes required us to reposition the microwave antenna, resulting in a slightly different excitation geometry. Rabi line scans were performed along the y direction (Fig.~\ref{fig:Cartoon}), cutting across the nanowire near the center, but away from the domain. The scans were obtained at various heights ranging over 94-284 nm to explore the z-dependence of the RF field localization effect.

\begin{figure} 
	\centering
	\includegraphics[width=\textwidth]{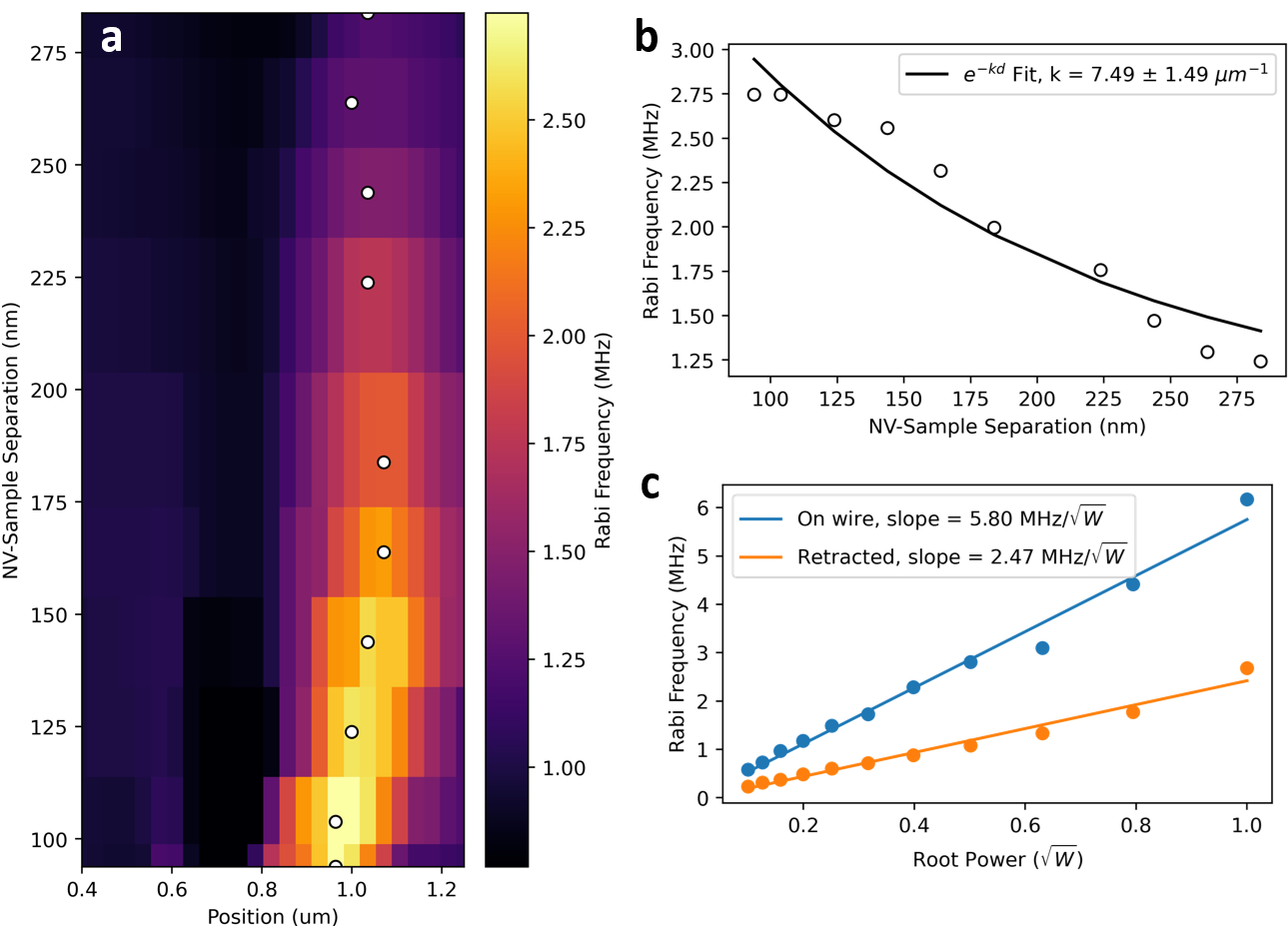}
	\caption{Height and power dependence of the flux-channeling effect. (a) Measurement of the Rabi frequency across the nanowire at different heights. The data is aligned using the topography measurements, with the edge positioned at 1 $\mu m$, but the maximum Rabi frequencies, which are marked by white dots, shift slightly from this point, likely because of drift between the scans. The microwave field modulation remains remarkably localized even at a few hundred nm distance. (b) Plot of the highest Rabi frequency points from (A) at each height and their accompanying exponential fit. (c) Measurement of the Rabi frequency at different applied microwave powers on with the tip both retracted off the sample and engaged over the enhanced side of the nanowire. Both results display an expected linear dependence on the root power, but the engaged measurement shows a 2.35$\times$ enhancement over the retracted measurement.}
	\label{fig:PnH} 
\end{figure}

The height dependence measurements, shown in Fig.~\ref{fig:PnH}(a), reveal a clear retention of enhanced Rabi frequency at larger separations, demonstrating the utility of this approach beyond scanning-tip NV systems. This flux-channeling-induced RF field localization makes the technique highly attractive for designing various device architectures, as it avoids the large spread of a conventional Oersted field. The shifting position of the RF contrast ``band'' with height was the result of drift of the scan head between the rapid, initial topography line scan and the much slower subsequent Rabi oscillation measurements at lower scan heights. The maximum values at various NV-sample separations, which are shown in Fig.~\ref{fig:PnH}(b), can be fitted with an exponential decay function. Using an $Ae^{-kd}$ fit, where $A$ is the Rabi frequency at the sample surface, $d$ is the NV-sample separation and $k$ is a decay constant, yields: $k = 7.49 \pm 1.49$ $\mu m^{-1}$, and $A = 4.07 \pm 0.56$ MHz. This result indicates that even our unoptimized wire design could provide over 4$\times$ increases in Rabi frequency or a 16$\times$ decrease in required microwave power with short NV-sample separations, with the effect remaining localized to within a few hundred nm vertically. While previous height-dependent studies of ferromagnetic dynamics have interpreted the decay constant $k$ as a magnon wavenumber\cite{Simon2022}, the $k$ here is merely an effective wavenumber - we are not resonantly exciting a specific mode nor do we expect this fit to be exact. We further discuss this point in the text below and under the Supporting Information with the aid of micromagnetic simulations.

To understand the power dependence, Rabi oscillations were measured manually, with the tip re-positioned and an ODMR scan performed in between the measurements to minimize the effects of tip-sample drift. At a site slightly shifted laterally from where the height line scans were performed and at 84 nm above the sample, we found a 2.35-fold enhancement of the external microwave field amplitude at the NV over the entire range of microwave powers (10 to 30 dBm) tested. Figure ~\ref{fig:PnH}(c) shows the Rabi frequency as a function of the applied microwave driving power for both the retracted tip and the tip engaged over the edge of the wire. The expected linear dependence with a 2.35-fold enhancement is seen for the tip engaged over the wire compared to the retracted tip. We conclude that flux-channeling induced localization of RF field applies over a large range of powers and remains highly localized both vertically and laterally.

\begin{figure} 
	\centering
	\includegraphics[width=\textwidth]{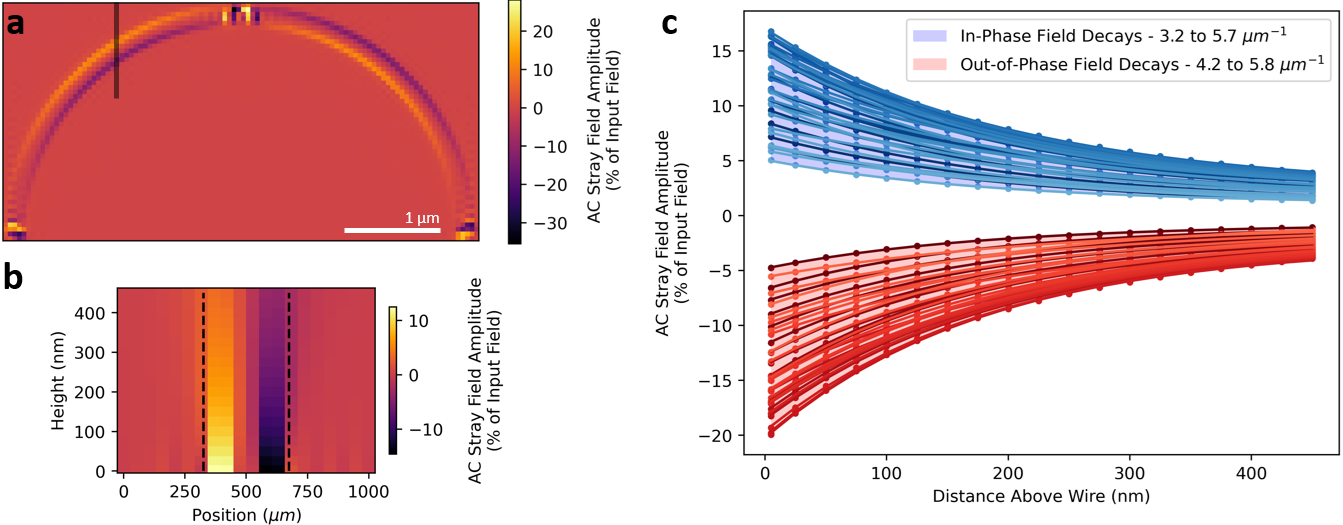}
	\caption{Micromagnetic simulations of flux-channeling in the measured nanowires. (a) Micromagnetic simulations of the nanowire’s contribution to the total microwave field along the orientation of the applied microwave field at 139 nm above the wire. The result is shown as a percentage of the applied field magnitude, with positive and negative values being the result of the magnet generating an in-phase field and a 180$^{\circ}$ phase-shifted field, respectively. (b) Simulated height dependence of the AC stray field, taken at the vertical slice marked in (A) by a translucent black line. The variation in microwave field amplitude remains localized to the edges of the nanowire, even at hundreds of nm above the surface. (c) Plots of the height dependence of the maximum field amplitudes (blue for in-phase, red for out-of-phase) at 40 separate vertical slices on the left side of the wire. The shading of the data corresponds to the x position of the slice, with darker lines denoting larger x, and lines correspond to each slice's $A e^{-kd}$ fits, where A is the amplitude, d is the height above the nanowire, and k is a decay constant. While every fit has a different decay constant k, they are all within 3.2 to 5.8 $\mu m^{-1}$.}
	\label{fig:sims} 
\end{figure}

We further analyzed our results using micromagnetic simulations (Fig.~\ref{fig:sims}). A sinusoidal field (0.1 mT, 2.85 GHz) was applied at a 45$^{\circ}$ angle in the YZ plane of the simulation window, and the magnetization of the wire was sampled and used to find the stray field over a 20 ns period. Outside the proximity of the domain wall pinning site, simulation results are the same whether or not the wire contains a domain wall, showing that our results are not dependent on any specific texture; see Supporting Information for details. Figure ~\ref{fig:sims}(a) shows the 2.85 GHz component of the stray field produced along the same 45$^{\circ}$ degree angle as the applied microwave field, normalized to the magnitude of the applied field. Positive values, shown with brighter colors, denote areas where the stray field is in-phase with the applied field, increasing the total amplitude via constructive interference, while the negative areas, shown with darker colors, denote areas where the stray field is 180$^{\circ}$ out-of-phase with the applied field, resulting in a net decrease of the total field amplitude via destructive interference. 

These results show an excellent qualitative match with our scans in Fig.~\ref{fig:scans}: both the bright and dark bands near the right and left edges of our wire are replicated, despite some variations in the exact values of the stray field. We do not expect quantitative agreement because of the uniform MW excitation used in our simulations, which is only an approximate representation of our MW antenna. Figure~\ref{fig:sims}(b) shows the height dependence of the microwave field modulation, taken across the dark grey cut in Fig~\ref{fig:sims}(a). Here, the microwave stray field is seen to remain remarkably localized to the edges of the wire, even at 450 nm above the wire, the highest height we sampled, matching the results shown in Fig.~\ref{fig:PnH}(a) well. Fig.~\ref{fig:sims}(c) shows the decay of the largest field amplitudes at an additional 40 cuts around the point slice shown in Fig.~\ref{fig:sims}(b), with the x position of the cut corresponding to the darkness of the points. These decays can all be fit to $A e^{-kd}$, like the experimental data, giving decay constants of 3 to 6 $\mu m^{-1}$, the upper end of which is within the fit error of our experimentally-determined value. This allows us to interpret this effective wavenumber using micromagnetic simulations, from which we can calculate a weighted average of the exponential decay as\begin{equation}
    S = \sum_{i}I(k_i)e^{-k_i t}
\end{equation}
where $I(k_i)$ is the absolute value of the spatial Fourier transform for spatial frequency $k_i$. This fitting yields an effective $k$ of 2.99 $\mu m^{-1}$, which falls slightly under the range of our simulated slices; see Supporting Information for details. We thus interpret our height dependence not as a single exponential decay, but as a distribution of decays arising from the range of spatial frequencies in our system, which our micromagnetic simulations show are primarily determined by the shape of the magnetic material. This suggests that the effective $k$ can be optimized by changing the shape of the structure. 

\begin{figure} 
	\centering
	\includegraphics[width=\textwidth]{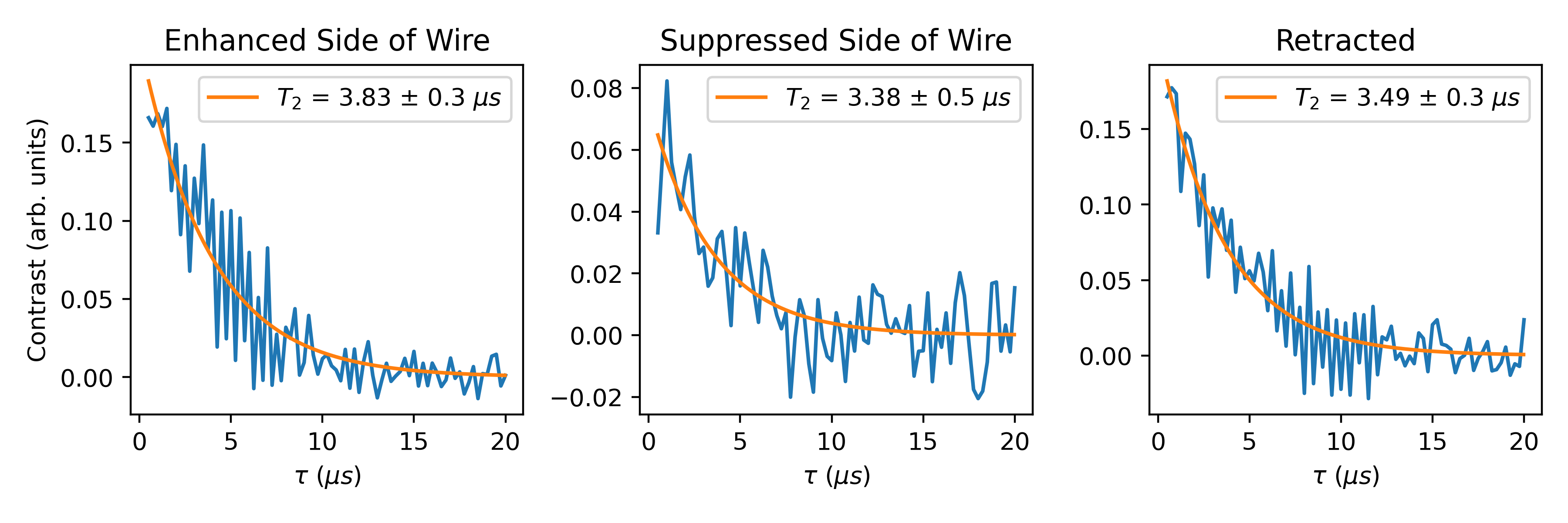}
	\caption{The results of spin echo measurements of the transverse coherence time, $T_2$, of the NV center over the enhanced section of the wire (94 ns Rabi period), over the suppressed section of the wire (288 ns Rabi period), and retracted from the surface (149 ns Rabi period). All values lie within 1 standard deviation of one another, and a small increase in coherence with shorter pulse times is not unexpected, as some decoherence will occur during qubit nutation.}
	\label{fig:T2} 
\end{figure}

Next, we turn to one of the conditions that must be satisfied for flux channeling to be useful for controlling qubits, namely, it should not significantly degrade the coherence time. It has been well-established that proximity to magnetic materials can dramatically alter relaxation timescale in NV centers\cite{Wolfe2014b,VanderSar2015,Du2017,Badea2023,yan_coherent_2024}; thus, we need to understand the local noise environment. Here, our use of the NV center as a probe stands out as a prototypical qubit that also allows us to measure decoherence behavior. To evaluate this, we performed spin-echo measurements of $T_2$ coherence, using the standard Hahn spin-echo sequence\cite{Hahn1950,Hanson2006}, over different sites on the second nanowire and compared to the results to those obtained for the retracted case. 

All measurements were restricted to five minutes because of small drifts in tip position, which, coupled with the spatial dependence of DC and AC magnetic environments, gradually shifted the NV $\ket{0} \rightarrow \ket{1}$ transition and Rabi frequencies in the on-sample measurements, leading to rapid accumulation of pulse errors. To increase measurement speed, a third diamond tip with a parabolic pillar (Qnami Quantilever MX+), which increases photon collection efficiency, was used. Fig.~\ref{fig:T2} shows the results of spin-echo measurements at three types of location. The (natural) $T_2$ time of the NV, obtained at a retracted position, was 3.49 µs, which increased to 3.83 $\mu s$ over the enhanced side of the wire and decreased to 3.38 $\mu s$ over the suppressed side. These values all lie within one standard deviation of one another, demonstrating that our flux channeling technique does not decrease $T_2$ and could indeed become a viable method for nanoscale control of qubits.

In addition to $T_2$ coherence, the effect of our nanowires on NV $T_1$ is also important to the viability of the technique. However, we could not measure localized $T_1$ times because of drift and measurement time constraints (full $T_1$ measurements can take hours). Regardless, operating below the FMR frequency of the ferromagnet implies that we are operating below the frequencies of the ferromagnet's thermal magnon gas, which should avoid the decrease in $T_1$ from magnon noise\cite{Du2017}, providing an advantage over magnon coupling schemes. Moreover, our Hahn spin echo measurements empirically show T2 (often the limiting factor in sensing experiments) is unaffected by the permalloy features.

We next discuss several potential approaches to optimizing these devices for microwave flux channeling, opening up new research areas in qubit control. (1) While permalloy is an excellent material for our proof-of-concept study, it is not commonly used in high-frequency magnetic cores due to eddy current generation. Because our MW field amplitudes are under 1 mT, saturation magnetization is not a primary concern in material choice as it would often be in conventional, macroscopic magnetic core design, suggesting that lower saturation, less conducting materials, such as Ni-Fe-Mo alloys or soft, electrically insulating ferrites, would be superior\cite{supermalloy_1948,mclyman_designing_1985}. (2) Geometry alterations could be made to target specific, narrower frequency ranges. While the high FMR frequencies of these wires ensured that we would be measuring in the regime where flux channeling enhances our MW fields, the real component of the permeability of a patterned ferromagnet peaks just below its FMR frequency\cite{El-Ghazaly2017}. Therefore, a smaller gap between the qubit transition frequencies and FMR frequency would allow for stronger Rabi enhancement. FMR frequency could be manipulated by modifying the geometry to increase or decrease the shape anisotropy. For example, a larger, straighter structure with lesser shape anisotropy will have a lower FMR frequency along its easy axis at low bias fields\cite{Rable2022}. However, as fields (and enhancement factors) become more localized, new integration strategies will be required to align these high-amplitude regions with the target qubit. Another technique reported for increasing anisotropy and, in turn, the FMR frequency is lamination\cite{el-ghazaly_increasing_2015}. (3) The efficiency could be improved by selecting feature thickness relative to the material's skin depth at the frequency of interest to reduce eddy currents while maximizing stray field amplitude. Finally, (4), decreasing the qubit-ferromagnet separation could offer a substantial increase in efficacy, as demonstrated by the exponential decay we observe.

In conclusion, we have demonstrated nanoscale imaging of flux channeling of a microwave magnetic field and position-dependent enhancement and suppression of qubit rotation. Unlike the previous approaches that involve manipulation of qubit dynamics with ferromagnets, flux channeling does not require an external magnetic field or specific magnetic texture and remains effective over a broad range of frequencies when the materials and nanopatterns are chosen properly. Furthermore, because we chose our geometry to ensure flux channeling would occur, rather than to maximize the strength of it, improvements should be possible through optimization of magnet geometries and material choice. Our technique could see potential applications where high amplitude microwave Oersted fields over a large spatial region are undesirable. Examples include: quantum sensing of biological samples, where heating from the microwave field could cause damage; spintronics, where local injection of microwaves may be desirable; creation of a soft-ferromagnet signal amplification scheme for the detection of weak signals, similar to that of \citeauthor{trifunovic_high-efficiency_2015}\cite{trifunovic_high-efficiency_2015}, but without the need to tune the ferromagnet’s FMR frequency to the target spin frequency; and finally, measurements of flux channeling with quantum sensors could prove useful in the development of nanoscale / mesoscopic transformers, where other techniques, such as spin-echo \cite{Maze2008} and quantum frequency mixing\cite{hu_nonlinear_2024,karlson_quantum_2024}, could be used to expand the range of measured frequencies.

\begin{acknowledgement}

\paragraph*{Funding:}
S.K. and A.B. acknowledged support provided by the National Science Foundation through the ExpandQISE award No. 2329067 and the Massachusetts Technology Collaborative through award number MTC-22032.
N.S. acknowledged support from the University of Chicago and the U.S. Department of Energy Office of Science National Quantum Information Science Research Centers (Q-NEXT).
\paragraph*{Author contributions:}
J.R. fabricated samples with J.D., performed the experimental measurements, ran the micromagnetic simulations, and analyzed the data with P.S. P.S., N.S., A.B., and S.K supervised the project. J.R. wrote the manuscript with substantial input from all coauthors. 
\paragraph*{Competing interests:}
Northeastern University and Pennsylvania State University are currently pursuing the IP associated with this work. The authors do not perceive any conflict of interest. 
\paragraph*{Data and materials availability:}
All data will be available upon reasonable request. 

\end{acknowledgement}

\begin{suppinfo}

Contains details of the experimental measurements, micromagnetic simulations, and diamond tip calibration; a description of the image processing performed along with the unprocessed data; additional micromagnetic simulation results

\end{suppinfo}

\bibliography{acs-achemso}

\end{document}